\documentclass[11pt]{article}

\newtheorem{theorem}{Theorem}[section]
\newtheorem{lemma}[theorem]{Lemma}

\newtheorem{claim}[theorem]{Claim}

\newtheorem{definition}[theorem]{Definition}

\newenvironment{proof}{\smallskip\noindent{\bf Proof:  }}
{\qquad\rule{2mm}{2mm}\smallskip}

\makeatletter
\def\eqalign#1{\,\vcenter{\openup\jot\m@th
  \ialign{\strut\hfil$\displaystyle{##}$&$\displaystyle{{}##}$\hfil
      \crcr#1\crcr}}\,}
\def\eqalignno#1{\displ@y \tabskip\@centering
  \halign to\displaywidth{\hfil$\displaystyle{##}$\tabskip\z@skip
    &$\displaystyle{{}##}$\hfil\tabskip\@centering
    &\llap{$##$}\tabskip\z@skip\crcr
    #1\crcr}}
\makeatother

\usepackage{epic}
\usepackage{eepic}
\usepackage{latexsym}

\newcommand{\Ot}{\tilde{O}}
\newcommand{\Omt}{\tilde{\Omega}}
\newcommand{\ignore}[1]{}
\newcommand{\tmpignore}[1]{}
\newcommand{\ee}{{\rm e}}
\newcommand{\reals}{\hbox{$\rlap{\rm I} \> \kern-.2mm{\rm R}$}}
\newcommand{\eq}{\;=\;}
\newcommand{\eps}{\epsilon}
\newcommand{\qu}{\quad}
\newcommand{\cT}{{\cal T}}

\newcommand{\combcolor}{{\sc Combined-Color}}

\title{Coloring $k$-colorable graphs using relatively small palettes \thanks{A
    preliminary version of this paper appeared in the proceedings of
    the 12th ACM-SIAM Symposium on Discrete Algorithms (SODA'01),
    Washington D.C., 2001, 
    pages 319--326.}}

\author{{\em Eran Halperin} ${}^\dagger$ \and
{\em Ram Nathaniel} ${}^\dagger$ \and {\em Uri Zwick}
\thanks{School of Computer Science,
Tel-Aviv University, Tel-Aviv 69978, Israel.
E-mail: {\tt \{heran,ramn,zwick\}}@cs.tau.ac.il.}}

\begin{document}
\maketitle

\begin{abstract}
\noindent
We obtain the following new coloring results:
\begin{itemize}
\item A 3-colorable graph on $n$ vertices with maximum degree~$\Delta$
  can be colored, in polynomial time, using $O((\Delta
  \log\Delta)^{{1}/{3}}\cdot\log{n})$ colors. This slightly improves
  an $O((\Delta^{{1}/{3}}\log^{1/2}\Delta)\cdot\log{n})$ bound given
  by Karger, Motwani and Sudan. More generally, $k$-colorable graphs
  with maximum degree $\Delta$ can be colored, in polynomial time, using
  $O((\Delta^{1-{2}/{k}}\log^{1/k}\Delta)\cdot\log{n})$ colors.
\item A 4-colorable graph on $n$ vertices can be colored, in
  polynomial time, using $\Ot(n^{7/19})$ colors. This improves an
  $\Ot(n^{2/5})$ bound given again by Karger, Motwani and Sudan.
  More generally, $k$-colorable graphs on $n$ vertices
  can be colored, in polynomial time, using $\Ot(n^{\alpha_k})$ colors, where
  $\alpha_5=97/207$, $\alpha_6=43/79$, $\alpha_7=1391/2315$,
  $\alpha_8=175/271$,  \ldots
\end{itemize}
The first result is obtained by a slightly more refined probabilistic
analysis of the semidefinite programming based coloring algorithm of
Karger, Motwani and Sudan. The second result is obtained by combining
the coloring algorithm of Karger, Motwani and Sudan, the combinatorial
coloring algorithms of Blum and an extension of a technique of Alon
and Kahale (which is based on the Karger, Motwani and Sudan algorithm)
for finding relatively large independent sets in graphs that are
guaranteed to have very large independent sets.
The extension of the Alon and Kahale result may be of independent interest.
\end{abstract}

\newcommand{\thealg}{
\begin{figure}[h!]
\begin{center}
\framebox{\hspace{0.6cm}\parbox{35pc}{
\vspace*{7pt}
{\bf Algorithm \combcolor:}\\[5pt]
{\bf Input:} A graph $G=(V,E)$ on $n$ vertices and an integer $k\ge
2$. \\
{\bf Output:} An $\Ot(n^{\alpha_k})$ coloring of~$G$, if $G$ is $k$-colorable.

\begin{enumerate}
\item If $k=2$, color the graph, in linear time, using 2 colors.

\item If $k=3$, use the algorithm of Blum and Karger \cite{BlKa97} to
  color the graph using $\Ot(n^{3/14})$ colors.

\item Assume, therefore, that $k\ge 4$. Repeatedly remove from the
  graph~$G$ vertices of degree less than $n^{\alpha_k/(1-2/k)}$. Let
  $U$ be the set of vertices so removed, and let $G[U]$ be the
  subgraph of $G$ induced by~$U$. Let $D$ be the {\em average\/}
  degree of~$G[U]$. It is easy to see that $D\le 2n^{\alpha_k/(1-2/k)}$.
\ignore{If $u,v\in U$, $(u,v)\in E$ and $u$
  was removed before $v$, then direct the edge $(u,v)$ from $u$ to
  $v$. The outdegree of each vertex of $G[U]$ is now at most
  $n^{\alpha_k/(1-2/k)}$.}

\item If $|U|\ge\frac{n}{2}$ then we can use the algorithm of Karger, Motwani and
  Sudan \cite{KaMoSu98} (Theorem~\ref{T-KMS}) to find an
  independent set of $G[U]$ of size $\Omt(n/D^{1-2/k})\ge
  \Omt(n^{1-\alpha_k})$, as $D\le 2n^{\alpha_k/(1-2/k)}$, and we have
  made progress of type~\ref{P-independent}.
\ignore{
color $G[U]$
  using $\Ot(\Delta^{1-2/k})\le \Ot(n^{\alpha_k})$ colors, as
  $\Delta\le n^{\alpha_k/(1-2/k)}$, and we have made progress of
  type~\ref{P-semicolor} (semicoloring).}

\item Otherwise, if $|U|<\frac{n}{2}$, let $W=V{-}U$. Note that
  $|W|\ge\frac{n}{2}$ and that the minimum degree $d_{\min}$ in~$G[W]$
  satisfies $d_{\min}\ge n^{\alpha_k/(1-2/k)}$.

\item For every $u,v\in W$ consider the set $S=N(u)\cap N(v)$. If
  $|S|\ge n^{(1-\alpha_k)/ (1-\alpha_{k-2}) }$, then apply the
  coloring algorithm recursively on $G[S]$ and $k-2$. If $G[S]$ is
  $(k-2)$-colorable, then the algorithm produces a coloring of $G[S]$
  using $\Ot(|S|^{\alpha_{k-2}})$ colors, from which an independent
  set of size $\Omt(|S|^{1-\alpha_{k-2}})\ge \Omt(n^{1-\alpha_k})$ is
  easily extracted, and we have made progress of
  type~\ref{P-independent}. If the coloring returned by the recursive
  call uses more than $\Ot(|S|^{\alpha_{k-2}})$ colors, we can infer
  that $G[S]$ is {\em not\/} $(k-2)$-colorable and thus, $u$ and $v$
  must be assigned the same color under any valid $k$-coloring of $G$,
  as we have made progress of type~\ref{P-samecolor}. 
  \ignore{We are tacitly assuming here that the coloring algorithm
    is deterministic so that it is guaranteed to produce a coloring
    using $\Ot(|S|^{\alpha_{k-2}})$ colors, if $G[S]$ is
    $(k-2)$-colorable. Our algorithm, however, is randomized. There
    are two ways of overcoming this difficulty. The first is to
    derandomize it using the technique of Mahajan and Ramesh
    \cite{MaRa99}. Alternatively, we can simply repeat the whole
    algorithm a sufficient number of times so that the error
    probability is small enough.}

\item Otherwise, we get that $|N(u)\cap N(v)|< n^{(1-\alpha_k)/
    (1-\alpha_{k-2}) }$, for every $u,v\in W$. Also, we know that the
  minimum degree in $G[W]$ is at least $d_{\min}\ge
  n^{\alpha_k/(1-2/k)}$.

\item We can now apply Blum's algorithm \cite{Blum94}
  (Theorem~\ref{T-Blum}), with $d_{\min}\ge n^{\alpha_k/(1-2/k)}$ and $s\le
  n^{(1-\alpha_k)/ (1-\alpha_{k-2}) }$, and obtain a collection $\cT$
  of $\Ot(n)$ subsets of~$W$ such that at least one $T\in \cT$
  satisfies $|T|\ge \Omt(\frac{d^2_{\min}}{s}) \ge
  \Omt\Bigl(n^{\frac{2\alpha_k}{(1-2/k)} -
    \frac{1-\alpha_k}{1-\alpha_{k-2}}}\Bigr)$, and~$T$ contains an
  independent set of size at least $(\frac{1}{k-1}-O(\frac{1}{\log
    n}))|T|$.

\item We now apply the extension of the Alon and Kahale \cite{AlKa98}
  technique (Theorem~\ref{T-Alon-Kahale-extension}) on $G[T]$, for
  each $T\in \cT$. In at least one of these runs we obtain an
  independent set of size $\Omt\Bigl(n^{\bigl(\frac{2\alpha_k}{1-2/k}
    - \frac{1-\alpha_k}{1-\alpha_{k-2}}\bigr)\cdot
    \frac{3}{k}}\Bigr)$. It is easy to check that
  $\bigl(\frac{2\alpha_k}{1-2/k} -
  \frac{1-\alpha_k}{1-\alpha_{k-2}}\bigr)\cdot \frac{3}{k} =
  1-\alpha_k$ (the sequence $\alpha_k$ is defined to satisfy this
  relation), so we have made progress of type~\ref{P-independent}.
\end{enumerate}
}\hspace{0.6cm}}
\end{center}
\caption{The new coloring algorithm.}
\label{F-rc}
\end{figure}
}

\section{Introduction}

\begin{table*}[t]
\renewcommand{\arraystretch}{2.15}
\newcommand{\dfrac}[2]{\displaystyle\frac{#1}{#2}}
\begin{center}
\begin{tabular}{||c|c|c|c|c|c|c||}\hline\hline
{\bf Coloring algorithm} & $k=3$ & $k=4$ & $k=5$ & $k=6$  & $k=7$ &
 $k=8$   \\
 \hline\hline
\Large
Wigderson & $\dfrac{1}{2}$ & $\dfrac{2}{3}$    & $\dfrac{3}{4}$     & $\dfrac{4}{5}$     & $\dfrac{5}{6}$    & $\dfrac{6}{7}$ \\
\cite{Wi83}     & 0.5 & 0.666 & 0.75    & 0.8     & 0.833 &
          0.857 \\ \hline
\Large
Blum      & $\dfrac{3}{8}$ &$\dfrac{ 3}{5}$    & $\dfrac{91}{131}$  &$\dfrac{105}{137}$  & $\dfrac{5301}{6581}$ & $\dfrac{10647}{12695}$ \\
\cite{Blum94}  & 0.375 & 0.6 & 0.694& 0.766 & 0.805 &
          0.838\\ \hline
Karger, Motwani, Sudan
   & $\dfrac{1}{4}$ & $\dfrac{2}{5}$    & $\dfrac{1}{2}$     & $\dfrac{4}{7}$     & $\dfrac{5}{8}$ & $\dfrac{2}{3}$ \\
\cite{KaMoSu98}   & 0.25& 0.4 & 0.5 & 0.571 & 0.625 & 0.666 \\
 \hline
\Large
Our Results & $\left[\dfrac{3}{14}\right]$ &
$\dfrac{7}{19}$ & $\dfrac{97}{207}$  &
$\dfrac{43}{79}$   & $\dfrac{1391}{2315}$ & $\dfrac{175}{271}$ \\
   $\star$  & (0.214) & 0.368  & 0.468 & 0.544 &
 0.600 & 0.645
          \\ \hline

\end{tabular}
\end{center}
\caption{\label{T-comparison} The exponents of the new coloring
 algorithms, and of the previously available algorithms, for $3\le
 k\le 8$. The $3/14$ exponent for $k=3$ is from Blum and Karger \protect\cite{BlKa97}.}
\end{table*}

Finding a 3-coloring of a given 3-colorable graph is a well known
NP-hard problem. Finding a 4-coloring of such a graph is also known to
be NP-hard (Khanna, Linial and Safra \cite{KhLiSa00} and Guruswami and
Khanna \cite{GuKh00}). Karger, Motwani and Sudan \cite{KaMoSu98} show,
on the other hand, using semidefinite programming, that a 3-colorable
graph on~$n$ vertices with maximum degree $\Delta$ can be colored, in
polynomial time, using
$O((\Delta^{{1}/{3}}\log^{1/2}\Delta)\cdot\log{n})$ colors.  Combining
this result with an old coloring algorithm of Wigderson \cite{Wi83}
they also obtain an algorithm for coloring arbitrary 3-colorable
graphs on~$n$ vertices using $O(n^{1/4}\log^{1/2}n)$ colors. By
combining the result of Karger {\em et al.\/} \cite{KaMoSu98} with a
coloring algorithm of Blum \cite{Blum94}, Blum and Karger
\cite{BlKa97} obtain a polynomial time algorithm that can color a
3-colorable graph using $\Ot(n^{3/14})$ colors.

The semidefinite programming based coloring algorithm of Karger,
Motwani and Sudan \cite{KaMoSu98} can also be used to color
$k$-colorable graphs of maximum degree $\Delta$ using
$\Ot(\Delta^{1-2/k})$ colors. Combined again with the technique of
Wigderson \cite{Wi83} this gives a polynomial time algorithm for
coloring $k$-colorable graph using $\Ot(n^{1-3/(k+1)})$ colors. Blum
\cite{Blum94} gives a combinatorial algorithm for coloring
$k$-coloring graphs using $\Ot(n^{\beta_k})$ color, where the
$\beta_k$'s satisfy a complicated recurrence relation. The first
values in the sequence are $\beta_3=\frac{3}{8}$,
$\beta_{4}=\frac{3}{5}$, $\beta_5=\frac{91}{131}$, \ldots The
algorithm of Karger {\em et al.\/} \cite{KaMoSu98} uses less colors than the
algorithm of Blum \cite{Blum94} for any $k\ge 3$. No combination of
the semidefinite programming based coloring algorithm of {\em et
  al.\/} \cite{KaMoSu98} with the combinatorial algorithm of Blum
\cite{Blum94} was given, prior to this work, for $k\ge 4$.

In this paper we present several improved coloring algorithms. Our
improvements fall into two different categories. We first consider the
semidefinite programming based coloring algorithm of Karger, Motwani
and Sudan \cite{KaMoSu98}. We show that the number of colors used by
this algorithm can be reduced, alas, by only a polylogarithmic factor.
Though the improvement obtained here is not very significant, we
believe that it is interesting as it is obtained not using tedious
calculations but rather using a simple refinement of the probabilistic
analysis given by Karger {\em et al.\/} \cite{KaMoSu98}. Furthermore, we can
show that this refined analysis is tight.

Having considered the algorithm of Karger {\em et al.\/}
\cite{KaMoSu98} on its own, we turn our attention to possible
combinations of that algorithm with the combinatorial algorithm of
Blum \cite{Blum94}. The $\Ot(n^{3/14})$ result of Blum and Karger
\cite{BlKa97} for $k=3$ is an example of such a combination. Although
no such combinations were previously reported for $k>3$, it is not
difficult to construct simple combinations of these algorithms that
would yield improved results. We go one step further and present
non-trivial combinations of these algorithms that yield even further
improvements.  In particular, our combinations use a third ingredient,
an extension of algorithm of Alon and Kahale \cite{AlKa98} that can be
used to find large independent sets in graphs that contain very large
independent sets. More specifically, Alon and Kahale \cite{AlKa98}
show that if a graph on $n$ vertices contains an independent set of
size $n/k+m$, for some fixed integer $k\ge 3$ and some $m>0$, then an
independent set of size $\Omt(m^{3/(k+1)})$ can be found in (random)
polynomial time. We extend this result and show that if a graph on $n$
vertices contains an independent set of size $n/\alpha$, where
$\alpha\ge 1$ is not necessarily integral, then an independent set of
size $\Omt(n^{f(\alpha)})$ can be found in (random) polynomial time,
where $f(\alpha)$ is a continuous function, described explicitly in
the sequel, that satisfies $f(k)=3/(k+1)$, for every integer $k\ge 2$.
This result may be of independent interest. Interestingly, the Alon
and Kahale \cite{AlKa98} result, and its extension, are based on the
algorithm of Karger, Motwani and Sudan \cite{KaMoSu98} that may also
be viewed as an algorithm for finding large independent sets.

Equipped with this new ingredient, we describe a combined coloring
algorithm that uses ideas from Blum \cite{Blum94}, Karger {\em et
  al.\/} \cite{KaMoSu98} and Alon and Kahale \cite{AlKa98} to color a
$k$-colorable graph using $\Ot(n^{\alpha_k})$ colors, where
$\alpha_4=7/19$, $\alpha_5=97/207$, $\alpha_6=43/79$,
$\alpha_7=1391/2315$, $\alpha_8=175/271$, $\ldots\;$ (See
Table~\ref{T-comparison} for a comparison of these bounds with the
previously available bounds.) An explicit, but complicated, recurrence
relation defining $\alpha_k$ for every $k$ is given later in the
paper. The new algorithm performs better than all the previously
available algorithms for $k\ge 4$. We obtain no improvement over the
$\Ot(n^{3/14})$ bound of Blum and Karger \cite{BlKa97} for $k=3$
(other than the polylogarithmic improvement mentioned earlier).

The rest of this paper is organized as follows. In Section~\ref{S-KMS}
we present our refinement to the\ignore{coloring} algorithm of
Karger\ignore{, Motwani and Sudan} {\em et al.\/} \cite{KaMoSu98}. In
  Section~\ref{S-Alon-Kahale} we present our extension of the
  technique of Alon and Kahale \cite{AlKa98}. In Section~\ref{S-Blum}
  we describe some coloring tools of Blum \cite{Blum94}. Finally, in
  Section~\ref{S-new} we describe our new coloring algorithm. We end
  in Section~\ref{S-concl} with some concluding remarks and open
  problems.

\section{A refinement analysis of the algorithm of Karger, Motwani}
\label{S-KMS}

Karger, Motwani and Sudan introduce the notion of a {\em vector
  coloring\/} of a graph, a notion that is closely related to
Lov\'{a}sz's {\em orthogonal representations\/} and to Lov\'{a}sz's
$\vartheta$-function (Lov\'{a}sz \cite{Lovasz79}, Gr\"{o}tschel {\em et
  al.}\/ \cite{GrLoSc93}):

\begin{definition}[\cite{KaMoSu98}]
  A {\em vector $\alpha$-coloring} of a graph $G=(V,E)$, where
  $V=\{1,2,\ldots,n\}$, is sequence of unit vectors
  $v_1,v_2,\ldots,v_n \in \reals^n$
  such that if $(i,j)\in E$,
  then $v_i\cdot v_j \le -\frac{1}{\alpha-1}$.
\end{definition}

It is easy to see that if $G$ is $k$-colorable then $G$ also has a
vector $k$-coloring. There are, however, graphs that are vector
$k$-colorable but are not $k$-colorable. A vector $k$-coloring of a graph
$G=(V,E)$, if one exists, can be found, in polynomial time, by solving
a semidefinite program. See \cite{KaMoSu98} for
details.\footnote{This statement is not completely accurate. What can
  be found in polynomial time is a vector $(k+\eps)$-coloring of the
  graph for, say, $\eps=2^{-n}$. The technical difficulties caused by
  this can be easily overcome. See \cite{KaMoSu98} for details.}
Karger, Motwani and Sudan \cite{KaMoSu98} also present the following
lemma which we use in Section~\ref{S-new}.

\begin{lemma}[\cite{KaMoSu98}]\label{L-alpha-1}
  Let $G=(V,E)$ be a vector $\alpha$-colorable graph, where
  $\alpha>2$.  Then, for every vertex $v\in V$, the subgraph of $G$
  induced by the neighbors of~$v$ is vector $(\alpha-1)$-colorable,
  and a vector $(\alpha-1)$-coloring of it can be found in polynomial
  time.
\end{lemma}

Karger {\em et al.}\/ \cite{KaMoSu98} show next that if $G=(V,E)$ is a
vector $k$-colorable graph on $n$ vertices with maximum
degree~$\Delta$, then an independent set of $G$ of size
$\Omega(\frac{n}{\Delta^{1-2/k}\log^{1/2}\Delta})$ can be found in
polynomial time. This easily implies that a vector $k$-colorable graph
on $n$ vertices with maximum degree $\Delta$ may be colored, in
polynomial time, using $O((\Delta^{1-2/k}\log^{1/2}\Delta)\cdot
\log n)$ colors. We obtain the following refinement of this result:

\begin{theorem}\label{T-KMS} Let $\alpha\ge 2$ and
  let $G=(V,E)$ be vector $\alpha$-colorable graph on $n$ vertices
  with average degree~$D$. Then, an independent set of $G$ of size
 at least $\Omega(\frac{n}{D^{1-2/\alpha}\log^{1/\alpha}D})$
can be found in polynomial time.
\end{theorem}

There are two minor differences and one more substantial difference
between Theorem~\ref{T-KMS} and the corresponding result of Karger
{\em et al.\/} \cite{KaMoSu98}. The first is that~$\alpha$ is not
assumed to be integral. The second is that the {\em maximum\/}
degree~$\Delta$ is replaced by the {\em average\/} degree~$D$. (The
$\Delta$ in the $\Omega((\Delta^{1-2/k}\log^{1/2}\Delta)\cdot \log n)$
bound {\em cannot\/} be replaced by~$D$, as the average degree, unlike
the maximum degree, may increase when vertices are removed from the
graph.) More interestingly, the exponent of $\log \Delta$ is reduced
from $1/2$ to $1/\alpha$, thus obtaining a poly-logarithmic improvement in
the number of colors needed to color low degree graphs. This
improvement, as we mentioned, is obtained using a simple modification
of the probabilistic argument of Karger {\em et al.\/}~\cite{KaMoSu98}.

We begin by presenting a proof of Theorem~\ref{T-KMS} for the case
$\alpha=3$. This allows us to explain the refined argument in the
simplest possible setting. We then explain the simple modifications
need to obtain a proof of the general case.

\begin{proof}\hspace*{-0.2cm}
  {\bf (of Theorem~\ref{T-KMS} for $\alpha=3$)}\quad \ignore{ Karger,
    Motwani and Sudan \cite{KaMoSu98} then describe a simple and
    elegant randomized algorithm for finding a large independent set
    in a vector 3-colorable graph $G=(V,E)$.} Let $v_1,v_2,\ldots,v_n$
  be a vector 3-coloring of~$G$. Let $D$ be the average degree of~$G$.
  Let $c=\sqrt{\frac{2}{3}\ln D-\frac{1}{3}\ln\ln D}$. (This is
  slightly different from the choice made by Karger {\em et al.\/}
  \cite{KaMoSu98}. They choose $c=\sqrt{\frac{2}{3}\ln D}$. It is
  the only change that we make to their algorithm.) Choose a random
  vector~$r$ according to the standard $n$-dimensional normal
  distribution. Let $I=\{i\in V \mid v_i\cdot r\ge c\}$. Let $n'=|I|$
  be size of $I$ and let $m'=|\{(i,j)\in E \mid i,j\in I\}|$ be the
  number of edges contained in $I$. An independent set $I'$ of size
  $n'-m'$ is then easily obtained by removing one vertex from each
  edge contained in $I$. We show that the expected size of~$I'$ is
  $\Omega(\frac{n}{(D\log D)^{1/3}})$.

\ignore{
Karger, Motwani and Sudan \cite{KaMoSu98} show that the expected size
of the independent set obtained is
$\Omega(\frac{n}{\Delta^{1/3}\log^{1/2}\Delta})$. Using a slightly
more refined analysis, and a slightly modified value of~$c$, we show
in the next section that the same algorithm actually produces an independent
set of size $\Omega(\frac{n}{(\Delta\log\Delta)^{1/3}})$. This easily
implies the improved coloring results claimed in the abstract and the
introduction.

We end this section by reviewing the analysis of Karger {\em et al.\/}
\cite{KaMoSu98}.}  Let $N(x)=\int_x^\infty \phi(y)dy$, where $\phi(x)=
\frac{1}{\sqrt{2 \pi}}\ee^{-\frac{x^2}{2}}$, denote the tail of the
standard normal distribution. It is well known that
$(\frac{1}{x}-\frac{1}{x^3})\phi(x) \le N(x) \le \frac{1}{x}\phi(x)$,
for every $x>0$. It is also known that if $v$ is an arbitrary unit
vector in $\reals^n$, and $r$ is a random vector chosen according to
the standard $n$-dimensional normal distribution, then the inner
product $v\cdot r$ is distributed according to the standard one
dimensional normal distribution. Furthermore, if $v_1$ and $v_2$ are
{\em orthogonal\/} unit vectors then the two random variables
$v_1\cdot r$ and $v_2\cdot r$ are {\em independent}.
It is easy to see,
then, that:
\begin{eqnarray*}
E[n'] &=& n \Pr[v_1 \cdot r \ge c]\;=\; nN(c) \;,\\
E[m'] &=& m \Pr[v_1 \cdot r \ge c \ {\rm and} \ v_2\cdot r \ge c
]\;,
\end{eqnarray*}
where $v_1$ and $v_2$ are two unit vectors such that $v_1\cdot v_2\le
-\frac{1}{2}$, and~$n$ and~$m$, respectively, are the number of
vertices and edges in the graph. It is not difficult to see that the
probability $\Pr[v_1 \cdot r \ge c \ {\rm and} \ v_2\cdot r \ge c ]$
is a monotone increasing function of the angle between $v_1$ and
$v_2$. As we would like to obtain an upper bound on the probability,
we may assume, therefore, that $v_1\cdot v_2=-\frac{1}{2}$.
Karger {\em et al.}\/ \cite{KaMoSu98} argue that
$$
    \Pr[v_1 \cdot r \ge c \ {\rm and} \ v_2 \cdot r \ge c]
 \;\le\;
      \Pr[(v_1+v_2) \cdot r \ge 2c) \;=\; N(2c)\;,$$
where the rightmost equality follows from the fact that $v_1+v_2$ is
also a unit vector. We obtain a slightly sharper upper bound on this
probability:

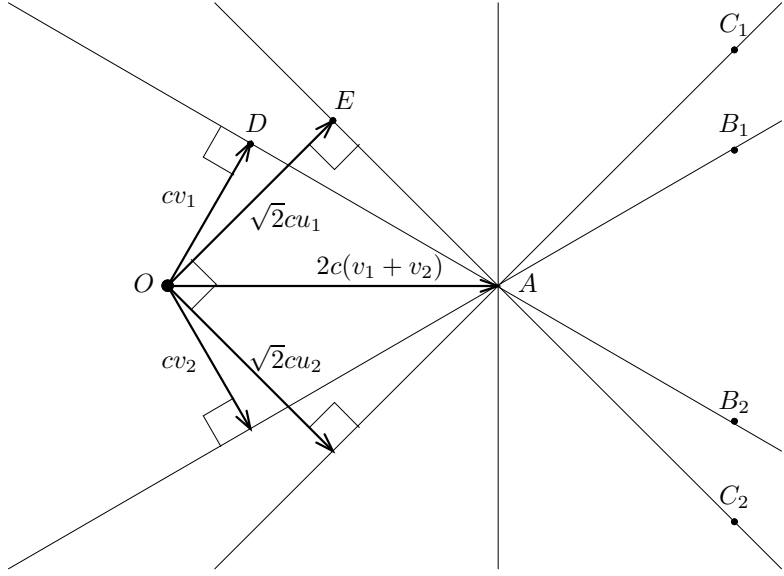
\begin{figure*}[t]
\begin{center}
\setlength{\unitlength}{0.00041667in}
\begingroup\makeatletter\ifx\SetFigFont\undefined%
\gdef\SetFigFont#1#2#3#4#5{%
  \reset@font\fontsize{#1}{#2pt}%
  \fontfamily{#3}\fontseries{#4}\fontshape{#5}%
  \selectfont}%
\fi\endgroup%
{\renewcommand{\dashlinestretch}{30}
\begin{picture}(9849,7239)(0,-10)
\put(2037,3612){\blacken\ellipse{150}{150}}
\put(2037,3612){\ellipse{150}{150}}
\put(9237,6612){\blacken\ellipse{80}{80}}
\put(9237,6612){\ellipse{80}{80}}
\put(9237,5337){\blacken\ellipse{80}{80}}
\put(9237,5337){\ellipse{80}{80}}
\put(9237,1887){\blacken\ellipse{80}{80}}
\put(9237,1887){\ellipse{80}{80}}
\put(9237,612){\blacken\ellipse{80}{80}}
\put(9237,612){\ellipse{80}{80}}
\put(4137,5712){\blacken\ellipse{80}{80}}
\put(4137,5712){\ellipse{80}{80}}
\put(3087,5412){\blacken\ellipse{80}{80}}
\put(3087,5412){\ellipse{80}{80}}
\path(12,12)(9837,5712)
\path(12,7212)(9837,1512)
\path(2637,12)(9837,7212)
\path(2637,7212)(9837,12)
\thicklines
\path(2037,3612)(3087,5431)
\path(3039.471,5198.640)(3087.000,5431.000)(2909.561,5273.630)
\path(2037,3612)(3087,1793)
\path(2909.561,1950.370)(3087.000,1793.000)(3039.471,2025.360)
\path(2037,3612)(4137,1512)
\path(3924.868,1618.066)(4137.000,1512.000)(4030.934,1724.132)
\thinlines
\path(2337,3948)(2655,3630)(2337,3312)
\path(3837,1812)(4155,2130)(4473,1812)
\path(4473,5412)(4155,5094)(3837,5412)
\path(2877,5022)(2487,5247)(2712,5637)
\path(2712,1572)(2487,1962)(2877,2187)
\path(6237,7212)(6237,12)
\thicklines
\path(2037,3612)(6237,3612)
\path(6012.000,3537.000)(6237.000,3612.000)(6012.000,3687.000)
\path(2037,3612)(4137,5712)
\path(4030.934,5499.868)(4137.000,5712.000)(3924.868,5605.934)
\put(2187,2562){\makebox(0,0)[b]{\smash{{{\SetFigFont{10}{12.0}{\rmdefault}{\mddefault}{\updefault}$cv_2$}}}}}
\put(4737,3762){\makebox(0,0)[b]{\smash{{{\SetFigFont{10}{12.0}{\rmdefault}{\mddefault}{\updefault}$2c(v_1+v_2)$}}}}}
\put(2187,4662){\makebox(0,0)[b]{\smash{{{\SetFigFont{10}{12.0}{\rmdefault}{\mddefault}{\updefault}$cv_1$}}}}}
\put(6612,3537){\makebox(0,0)[b]{\smash{{{\SetFigFont{10}{12.0}{\rmdefault}{\mddefault}{\updefault}$A$}}}}}
\put(9237,837){\makebox(0,0)[b]{\smash{{{\SetFigFont{10}{12.0}{\rmdefault}{\mddefault}{\updefault}$C_2$}}}}}
\put(9237,2037){\makebox(0,0)[b]{\smash{{{\SetFigFont{10}{12.0}{\rmdefault}{\mddefault}{\updefault}$B_2$}}}}}
\put(9237,5562){\makebox(0,0)[b]{\smash{{{\SetFigFont{10}{12.0}{\rmdefault}{\mddefault}{\updefault}$B_1$}}}}}
\put(9237,6837){\makebox(0,0)[b]{\smash{{{\SetFigFont{10}{12.0}{\rmdefault}{\mddefault}{\updefault}$C_1$}}}}}
\put(1737,3537){\makebox(0,0)[b]{\smash{{{\SetFigFont{10}{12.0}{\rmdefault}{\mddefault}{\updefault}$O$}}}}}
\put(3537,4362){\makebox(0,0)[b]{\smash{{{\SetFigFont{10}{12.0}{\rmdefault}{\mddefault}{\updefault}$\sqrt{2}cu_1$}}}}}
\put(3537,2562){\makebox(0,0)[b]{\smash{{{\SetFigFont{10}{12.0}{\rmdefault}{\mddefault}{\updefault}$\sqrt{2}cu_2$}}}}}
\put(4287,5862){\makebox(0,0)[b]{\smash{{{\SetFigFont{10}{12.0}{\rmdefault}{\mddefault}{\updefault}$E$}}}}}
\put(3162,5562){\makebox(0,0)[b]{\smash{{{\SetFigFont{10}{12.0}{\rmdefault}{\mddefault}{\updefault}$D$}}}}}
\end{picture}
}
\end{center}
\caption{Upper bounding $\Pr[v_1 \cdot r \ge c \ {\rm and} \ v_2\cdot r \ge
  c]$ when $||v_1||=||v_2||=1$ and $v_1\cdot
  v_2=-\frac{1}{2}$.}
\label{F-draw}
\end{figure*}

\begin{claim}\label{claim}
  If $v_1$ and $v_2$ are unit vectors such that $v_1\cdot v_2 =
  -\frac{1}{2}$ then
$$\Pr[v_1 \cdot r \ge c \ {\rm and} \ v_2\cdot r \ge c ] \;\le\;
N(\sqrt{2}c)^2 \;.$$
\end{claim}

\begin{proof}
  Let $v_1$ and $v_2$ be two unit vectors such that $v_1\cdot
  v_2=-\frac{1}{2}$. Note that $v_1$ and $v_2$ form an angle
  of~$120^\circ$. Let $D$ be the tip of $cv_1$. Draw a line
  perpendicular to $cv_1$ that passes through $D$. Similarly, draw a
  line perpendicular to $cv_2$ that passes through the tip of $cv_2$, as
  shown in Figure~\ref{F-draw}. It is easy to see that these two lines
  intersect at the point $A$ which is $2c(v_1+v_2)$. (This follows
  from the fact that $\angle DOA=60^\circ$ so that $\angle
  DAO=30^\circ$ and the fact that $\sin 30^\circ = \frac{1}{2}$.
  Note that $v_1+v_2$ is also a unit vector.)
  The projection of a standard $n$-dimensional normal vector $r$ on
  the plane spanned by $v_1$ and $v_2$ is a standard 2-dimensional
  normal vector which we denote by~$r'$. Note that $v_1\cdot r=v_1\cdot
  r'$ and $v_2\cdot r = v_2\cdot r'$. The probability that we have to
  bound is therefore the probability that the random vector $r'$
  falls into the wedge defined by the angle $\angle B_1AB_2$. Karger,
  Motwani and Sudan \cite{KaMoSu98} bound this probability by the
  probability that~$r'$ falls to the right of the vertical line that
  passes through $A$, which is $N(2c)$.

  Let $u_1$ and $u_2$ be unit vectors in the plane spanned by $v_1$
  and $v_2$ such that the angle formed by them and $v_1+v_2$ is
  $45^\circ$ (see Figure~\ref{F-draw}). Draw a line through $A$ which
  is perpendicular to $u_1$. Similarly, draw a line through $A$ which
  is perpendicular to~$u_2$. Let $E$ be the point on the first line in
  the direction of $u_1$. A simple calculation shows that
  $OE=\sqrt{2}c$. We bound the probability that $r'$ falls into the
  wedge formed by $\angle B_1AB_2$ by the probability that it falls
  into the wedge formed by $\angle C_1AC_2$. This probability is just
  $\Pr[u_1\cdot r'\ge \sqrt{2}c \ {\rm and} \ u_2\cdot r'\ge
  \sqrt{2}c]$. As $u_1\cdot u_2=0$, the events $u_1\cdot r'\ge
  \sqrt{2}c$ and $u_2\cdot r'\ge \sqrt{2}c$ are {\em independent}.
  Thus, this probability is just $N(\sqrt{2}c)^2$.
\end{proof}

Using a more complicated analysis, presented in Appendix~\ref{A-lower},
we can show that $\Pr[v_1 \cdot r \ge c \ {\rm and} \ v_2\cdot r
\ge c ] = \Omega(\frac{1}{c^2}\ee^{-2c^2})$. Thus, the bound given in
Claim~\ref{claim} is asymptotically tight.

We are now back in the proof of Theorem~\ref{T-KMS}. As $m\le nD/2$, we
get that
$$
E[n'-m'] \;\ge\; nN(c)-\frac{n D }{2}N(\sqrt{2}c)^2 \;=\;
n\left(N(c)-\frac{ D }{2}N(\sqrt{2}c)^2\right)\;.$$
  With
  $c=\sqrt{\frac{2}{3}\ln D - \frac{1}{3}\ln\ln D }$ we have
  $\ee^{3c^2/2}=\frac{D}{\ln^{1/2}D}$ and therefore
$$\frac{N(c)}{N(\sqrt{2}c)^2} >
\frac{(\frac{1}{c}-\frac{1}{c^3})\frac{1}{\sqrt{2\pi}}\ee^{-c^2/2}}
{\frac{1}{2c^2}\frac{1}{2\pi}\ee^{-2c^2}} >
\sqrt{2\pi}\cdot c\,\ee^{3c^2/2} > D\;.$$
Thus,
$$E[n'-m'] \ge \frac{n}{2}
(\frac{1}{c}-\frac{1}{c^3}) \frac{1}{\sqrt{2\pi}} \ee^{-c^2/2} =
\Omega( \frac{n}{(D\ln D)^{1/3}})\;,$$
and the proof of the theorem (for $\alpha=3$) is completed.
\end{proof}

The proof of the theorem for general $\alpha$ is very similar. We choose
$$c=\sqrt{(1-\frac{2}{\alpha})(2\ln D - \ln\ln D)}\;.$$ It is then not
difficult to see that the probability $\Pr[v_1 \cdot r \ge c \ {\rm
  and} \ v_2\cdot r \ge c ]$, when $v_1\cdot v_2=-\frac{1}{\alpha-1}$,
is upper bounded by $N(\sqrt{\frac{k-1}{k-2}}\,c)^2$, and the expected
size of the independent set $I'$ is indeed
$\Omega(\frac{n}{D^{1-2/\alpha}\log^{1/\alpha}D})$.

\ignore{
 \cite{KaMoSu98} get that
$$\frac{N(c)}{N(2c)} \;>\; \frac{
  (\frac{1}{c}-\frac{1}{c^3})\frac{1}{\sqrt{2\pi}} \ee^{-c^2/2}}
{ \frac{1}{2c} \frac{1}{\sqrt{2\pi}} \ee^{-2c^2}} \;=\;
2(1-\frac{1}{c^2})\ee^{3c^2/2} \;\ge\;  D \;,$$
where in the last inequality we assume that $\ln D \ge 3$ so that
$c^2\ge 2$. Thus,
$$E[n'-m'] \;\ge\; \frac{n}{2}N(c) \;\ge\; \frac{n}{2}
(\frac{1}{c}-\frac{1}{c^3}) \frac{1}{\sqrt{2\pi}} \ee^{-c^2/2} \;=\;
\Omega( \frac{n}{ D ^{1/3}\log^{1/2} D })\;.$$

\medskip We now use the algorithm of Karger {\em et al.\/}
\cite{KaMoSu98} with $c=\sqrt{\frac{2}{3}\ln\Delta -
  \frac{1}{3}\ln\ln\Delta}$, so that $\ee^{3c^2/2} =
\frac{\Delta}{\ln^{1/2}\Delta}$.  It is easy to verify that
$$\frac{N(c)}{N(\sqrt{2}c)^2} \;>\;
\frac{(\frac{1}{c}-\frac{1}{c^3})\frac{1}{\sqrt{2\pi}}\ee^{-c^2/2}}
{\frac{1}{2c^2}\frac{1}{2\pi}\ee^{-2c^2}} \;>\;
\sqrt{2\pi}\cdot c\,\ee^{3c^2/2} \;>\; \Delta\;.$$
Thus, with the notations of the previous section we get that
$$E[n'-m'] \;\ge\; \frac{n}{2}
(\frac{1}{c}-\frac{1}{c^3}) \frac{1}{\sqrt{2\pi}} \ee^{-c^2/2} \;=\;
\Omega( \frac{n}{(\Delta\ln\Delta)^{1/3}})\;.$$
This implies the improvement claimed.
}

\section{The Alon-Kahale algorithm and its extension}
\label{S-Alon-Kahale}

Alon and Kahale \cite{AlKa98} obtained the following result:

\begin{theorem}\label{T-Alon-Kahale}
Let $G=(V,E)$ be a graph on $n$ vertices that contains an
independent set of size at least $\frac{n}{k}+m$,
where $k \geq 3$ is an integer. Then, an independent set of $G$ of
size $\tilde{\Omega}(m^{3/(k+1)})$ can be found in polynomial time.
\end{theorem}

Here we prove the following extension of their result:

\begin{theorem}\label{T-Alon-Kahale-extension}
Let $G=(V,E)$ be a graph on $n$ vertices that contains an
independent set of size at least~$\frac{n}{\alpha}$, where
$\alpha\ge 1$. Let $k=\lfloor \alpha \rfloor$.
Then, an independent set of $G$ of
size $\tilde{\Omega}(n^{f(\alpha)})$ can be found in polynomial time,
where
$$f(\alpha) \eq \frac{\alpha(\alpha-1)}{ k\left(\alpha(\alpha-k) +
    \frac{(k-1)(k+1)}{3}\right) }\;.$$
In particular, $f(\alpha)=1$,
if $1\le \alpha\le 2$, $f(\alpha)=\frac{\alpha}{2(\alpha-1)}$, if
$2\le \alpha\le 3$, and $f(k)=\frac{3}{k+1}$ for every integer $k\ge 1$.
Also, the function $f(\alpha)$ satisfies the
functional equation $f(\alpha)=1\left/ \left( 1+
    \frac{1-{2}/{\alpha}}{f(\alpha-1)} \right) \right.$, for every
$\alpha\ge 2$.
\end{theorem}
\ignore{
\begin{theorem}\label{T-Alon-Kahale-extension}
Let $G=(V,E)$ be a graph on $n$ vertices that contains an
independent set of size at least~$\frac{n}{\alpha}$, where
$\alpha=k+y$, $k\ge 2$ is an integer and $0\le y<1$.
Then, an independent set of $G$ of
size $\tilde{\Omega}(n^{a(k,y)})$ can be found in polynomial time,
where
$$a(k,y)\eq \frac{(k+y)(k+y-1)}{2(y+1)^2 + \sum_{i=3}^k (k+i)(k+i-1)}
\eq \frac{(k+y)(k+y-1)}{ k(y^2+ky + \frac{(k-1)(k+1)}{3}) }\;.$$
In particular, $a(k,0)=\frac{3}{k+1}$ and $a(2,y)=\frac{2+y}{2(1+y)}$.
\end{theorem}
}

We only use this result for $\alpha=k+O(\frac{1}{\log n})$, where
$k\ge 2$ is an integer. As $f(k+O(\frac{1}{\log
  n}))=\frac{3}{k+1}+O(\frac{1}{\log n})$, we still get in this case
an independent set of size $\Omt(n^{3/(k+1)})$. For completeness, we
give a proof of the more general result. The proof of
Theorem~\ref{T-Alon-Kahale-extension} follows from the following two lemmas:

\ignore{ To prove the theorem, we use the following two lemmas of
  Karger Motwani and Sudan \cite{KaMoSu98}:

\begin{lemma}\label{deltaterms}\label{L-delta}
  Let $G=(V,E)$ be a vector $\alpha$-colorable graph on $n$ vertices,
  where $\alpha \geq 2$. Let $\Delta$ be the maximum degree
  in~$G$.  Then, an independent set of~$G$ of size
  $\tilde{\Omega}(\frac{n}{\Delta^{1-2/\alpha}})$ can be found in
  polynomial time.
\end{lemma}

\begin{lemma}\label{neighbours}\label{L-alpha-1}
Let $G=(V,E)$ be a vector $\alpha$-colorable graph, where $\alpha>2$.
Then, for every vertex $v\in V$, a vector $(\alpha-1)$-coloring of the
subgraph of $G$ induced by the neighbors of~$v$ can be found in
polynomial time.
\end{lemma}

As pointed out in Karger {\em et al.\/} \cite{KaMoSu98}, a graph is
vector 2-colorable if and only if it is 2-colorable. It is easy to see
that a graph is vector $\alpha$-colorable, with $\alpha<2$, if and
only if the graph contains no edges. Thus, it follows from
Lemma~\ref{L-alpha-1} that if $G=(V,E)$ is vector $\alpha$-colorable,
where $2\le \alpha < 3$, then the neighbors of each vertex in~$G$ form
an independent set.  We also need the following lemma: }

\begin{lemma}\label{vec-coloring}
Let $G = (V,E)$ be a graph on $n$ vertices with
an independent set of size at least $\frac{n}{\alpha}$, where
$\alpha \geq 2$. Then, a subset $S\subseteq V$ of size $|S|\ge
\frac{n}{\log n}$, and a vector $(\alpha+O(\frac{1}{\log
  n}))$-coloring of $G[S]$, the subgraph of $G$ induced by~$S$, can be
found in polynomial time.
\end{lemma}

\begin{proof}
Assume that
  $V=\{1,2,\ldots,n\}$ and consider the natural semidefinite
  programming relaxation of the maximum independent set problem:
$$\begin{array}{lc}
{\rm Maximize} & \displaystyle\sum_{i = 1}^n \frac{1+v_0\cdot v_i}{2} \\[10pt]
{\rm s.t.} & (v_0+v_i)\cdot (v_0 + v_j) = 0 \;,\; (i,j) \in E \\
&\|v_i\| = 1 \;,\; 1\le i\le n
\end{array}$$
An almost optimal solution $v_0,v_1,\ldots,v_n$ of this
semidefinite program can be found in polynomial time. As $G$ is
assumed to contain an independent set of size at least $n/\alpha$, we
may assume that
$$
\sum_{i=1}^n v_0\cdot v_i \;\geq\; \Bigl(\frac{2}{\alpha} - 1 -
\frac{1}{\log n}\Bigr)n \;.$$
(The $-1/\log n$ term comes from the fact
that $v_0,v_1,\ldots,v_n$ is only an almost optimal solution of the
program. We can make this term much smaller if we wish, but $1/\log n$ is
small enough for our purposes.) We now use the following simple facts:

\begin{claim}\label{C-1}
  If $\sum_{i=1}^n x_i \ge \gamma n$ and $x_i\le 1$, for every $1\le
  i\le n$,  then for any
  $\eps\ge 0$, at least $\eps n$ of the $x_i$'s satisfy
  $x_i>(\gamma-\eps)/(1-\eps)$.
\end{claim}

Indeed, if the claim is not satisfied then $ \sum_{i=1}^n x_i <
(1-\eps)n\cdot (\gamma-\eps)/(1-\eps) + \eps n = \gamma n$,
a contradiction. It is easy to check that
$(\gamma-\eps)/(1-\eps)>\gamma-2\eps$, if $\eps<(1+\gamma)/2$.
 Using this fact with $x_i=v_0\cdot v_i$,
$\gamma=\frac{2}{\alpha} - 1 - \frac{1}{\log n}$, and
$\eps=\frac{1}{\log n}$, we get that for at least $\frac{n}{\log n}$
of the vectors satisfy $v_0\cdot v_i > \frac{2}{\alpha} - 1 -
\frac{3}{\log n}$. Thus, if
$S=\{\,1\le i\le n\mid v_0\cdot v_i > \frac{2}{\alpha}-1-\frac{3}{\log
  n}\}$, then $|S|\ge \frac{n}{\log n}$.


\begin{claim}\label{C-2}
  Let $v_0,v_i$ and $v_j$ be unit vectors such that $v_i\ne v_0$,
  $v_j\ne v_0$ and
  $(v_0+v_i)\cdot(v_0+v_j)=0$. Let $v'_i$ and $v'_j$, respectively,
  be the normalized projections of $v_i$ and $v_j$ on the space
  orthogonal to~$v_0$. Then
$$v'_i\cdot v'_j \eq - \sqrt{\frac{1+(v_0\cdot v_i)}{1-(v_0\cdot v_i)}}
\cdot \sqrt{\frac{1+(v_0\cdot v_j)}{1-(v_0\cdot v_j)}} \;.$$
\end{claim}

\ignore{
Indeed,
$$v'_i \eq
\frac{v_i-(v_0\cdot v_i)v_0}{||v_i-(v_0\cdot v_i)v_0||}
\eq \frac{v_i-(v_0\cdot v_i)v_0}{\sqrt{(v_i-(v_0\cdot v_i)v_0)\cdot
    (v_i-(v_0\cdot v_i)v_0)}} \eq \frac{ v_i-(v_0\cdot v_i)v_0
  }{\sqrt{1-(v_0\cdot v_i)^2}}\;.$$
(Recall that $v_i$ is a unit vector so $v_i\cdot v_i=1$.) As
$(v_0+v_j)\cdot(v_0+v_i)=0$, we get that
that $v_i\cdot v_j = -1-v_0\cdot v_i-v_0\cdot v_j$.
Thus,
$$v'_i\cdot v'_j = \frac{(v_i-(v_0\cdot v_i)v_0)\cdot (v_j-(v_0\cdot
  v_j)v_0)}
{\sqrt{(1-(v_0\cdot v_i)^2)(1-(v_0\cdot v_j)^2)}}\;.
$$
The numerator of this expression can now be simplified as follows:
\begin{eqnarray*}
(v_i-(v_0\cdot v_i)v_0)\cdot (v_j-(v_0\cdot
  v_j)v_0) &=& v_i\cdot v_j - (v_0\cdot v_i)(v_0\cdot v_j) \\
&=&
-1 - (v_0\cdot v_i) - (v_0\cdot v_j) - (v_0\cdot v_i)(v_0\cdot v_j) \\
&=& -(1+(v_0\cdot v_i))(1+(v_0\cdot v_j))\;,
\end{eqnarray*}
and the claim follows.
}

\begin{proof}
Let $a_i=v_0\cdot v_i$ and $a_j=v_0\cdot v_j$. Then
$$\begin{array}{c}
\displaystyle v'_i \eq
\frac{v_i-a_iv_0}{||v_i-a_iv_0||} \\[10pt]
\displaystyle \eq \frac{v_i-a_iv_0}{\sqrt{(v_i-a_iv_0)\cdot
    (v_i-a_iv_0)}} \eq \frac{ v_i-a_iv_0
  }{\sqrt{1-a_i^2}}\;.\end{array}$$
(Recall that $v_i$ is a unit vector so $v_i\cdot v_i=1$.) .
Thus,
$$v'_i\cdot v'_j = \frac{(v_i-a_iv_0)\cdot (v_j-a_j v_0)}
{\sqrt{(1-a_i^2)(1-a_j^2)}}\;.
$$
As $(v_0+v_j)\cdot(v_0+v_i)=0$, we get that $v_i\cdot v_j \eq
-1 - v_0\cdot v_i - v_0\cdot v_j \eq -1-a_i-a_j$, and the numerator of
the expression given above for $v_i'\cdot v_j'$ can be simplified as
follows:
$$\begin{array}{c}
(v_i-a_iv_0)\cdot (v_j-a_jv_0) \eq v_i\cdot v_j - a_ia_j \\[3pt]
\eq
-1 - a_i - a_j - a_ia_j
\eq -(1+a_i)(1+a_j)\;,
\end{array}$$
and the claim follows.
\end{proof}

\medskip We continue now with the proof of Lemma~\ref{vec-coloring}.
Recall that $S=\{i\mid v_0\cdot v_i > \beta\}$, where $\beta =
\frac{2}{\alpha}-1-\frac{2}{\log n}$, and that $|S|\ge \frac{n}{\log
  n}$.  We may assume that $v_i\ne v_0$, for every $i\in
S$. Otherwise, we can very slightly perturb $v_0$. (Recall that the
vectors $v_0,v_1,\ldots,v_n$ form, in any case, only an almost optimal
solution of the semidefinite program.)
Suppose now that $i,j\in S$ and $(i,j)\in E$. Thus $v_0\cdot
v_i>\beta$, $v_0\cdot v_j > \beta$ and $(v_0+v_i)\cdot(v_0+v_j)=0$.
Let $v'_i$ and $v'_j$ be the normalized projections of $v_i$ and $v_j$
on the space orthogonal to~$v_0$. The expression given for $v'_i\cdot
v'_j$ in Claim~\ref{C-2} is {\em decreasing\/} in both $v_0\cdot v_i$
and $v_0\cdot v_j$. Thus,
$$v'_i\cdot v'_j \;\le\; -\frac{1+\beta}{1-\beta} \eq
-\frac{1}{\alpha-1+O(\frac{1}{\log n})}\;.$$
We obtained, therefore, a  vector $(\alpha+ O(\frac{1}{\log
n}))$-coloring of $G[S]$. This completes the proof.
\end{proof}

\begin{lemma}\label{L-2}
  Let $\alpha\ge 1$, and let $G=(V,E)$ a vector $\alpha$-colorable
  graph on $n$ vertices. Then, an independent set of~$G$ of size
  $\Omt(n^{f(\alpha)})$ can be found in polynomial time, where
  $f(\alpha)$ is as in Theorem~\ref{T-Alon-Kahale-extension}.
\end{lemma}

\begin{proof}
  The proof is by induction on $k=\lfloor \alpha \rfloor$.  Assume at
  first that $k=1$. It is easy to see that a graph is vector
  $\alpha$-colorable, for some $\alpha<2$, if and only if the graph
  contains no edges. Thus, $V$ is an independent set of size~$n$.

  Assume, therefore, that $k\ge 2$. Let $\Delta$ be the maximum degree
  of~$G$. We describe two ways of finding independent sets of $G$.
  Using the algorithm of Karger, Motwani and Sudan \cite{KaMoSu98}
  (Theorem~\ref{T-KMS}), we can find, in polynomial time, an
  independent set of $G$ of size $\Omt({n}/{\Delta^{1-2/\alpha}})$.
  Alternatively, let $v$ be a vertex of~$G$ of degree~$\Delta$ and let
  $N(v)$ be the set of its neighbors. It follows from
  Lemma~\ref{L-alpha-1} that the subgraph $G[N(v)]$ induced by~$N(v)$
  is vector $(\alpha-1)$-colorable. By the induction hypothesis, we
  can find in $G[N(v)]$, in polynomial time, an independent set of
  size $\Omt(\Delta^{f(\alpha-1)})$. This independent set is also an
  independent set of~$G$.  Taking the larger of these two independent
  sets, we obtain an independent set of $G$ of size
  $$
\displaystyle\Omt(\max\{\; \frac{n}{\Delta^{1-2/\alpha}} \;,\;
  \Delta^{f(\alpha-1)} \;\})
\displaystyle\;\ge\; \Omt( n^{1\left/ \left( 1+
        \frac{1-{2}/{\alpha}}{f(\alpha-1)} \right) \right.} ) \eq
  \Omt(n^{f(\alpha)})\;,
  $$
  as required. It is easy to verify, by induction,\ignore{using
    the relation $f(\alpha) = 1/ ( 1+
    \frac{1-{2}/{\alpha}}{f(\alpha-1)} ) $} that $f(\alpha) =
  \frac{\alpha(\alpha-1)}{ k\left(\alpha(\alpha-k) +
      \frac{(k-1)(k+1)}{3}\right) }$, where $k=\lfloor \alpha\rfloor$.
  We omit the straightforward details. This completes the proof of the lemma.
\end{proof}

\ignore{
As $f(\alpha)= \frac{\alpha}{2(\alpha-1)}$, for $2\le \alpha\le 3$,
this completes the proof for $k=2$.

Assume therefore that $k>2$. As before, we get can find, in polynomial
time, a subset $S\subseteq V$ such that $|S|\ge \frac{n}{\log n}$ and
a vector $\alpha'$-coloring of $G[S]$, where $\alpha'=\alpha+
O(\frac{1}{\log n})$. Let~$\Delta$ be the maximum degree $G[S]$. We
again find two independent sets of~$G[S]$. The first, obtained using the
algorithm of Karger {\em et al.\/} \cite{KaMoSu98} is of size
$\Omt({n}/{\Delta^{1-2/\alpha}})$. To obtain the second independent
set, we find a vertex $v$ of $G[S]$ of degree~$\Delta$, and
we find a vector $(\alpha'-1)$-coloring of $N(v)\cap S$. By running
the algorithm of Karger {\em et al.\/} \cite{KaMoSu98} on $G[N(v)\cap
S]$, we obtain an independent

 where $\Delta$ is the maximum
degree in~$G[S]$. Alternatively, if we consider the neighborhood
$N(v)$ of a vertex $v\in S$ of degree~$\Delta$, we get a subgraph
of~$G$ of size~$\Delta$ that is $(\alpha'-1)$-colorable. By the
induction hypothesis, we can find in this subgraph an independent set
of size

\end{proof}
}

\medskip
We now present a proof of Theorem~\ref{T-Alon-Kahale-extension}.

\begin{proof}
  {\bf (of Theorem~\ref{T-Alon-Kahale-extension})}\quad Suppose that
  $G=(V,E)$ contains an independent set of size $n/\alpha$. By
  Lemma~\ref{vec-coloring}, we can find, in polynomial time, a subset
  $S\subseteq V$ of size $|S|\ge \frac{n}{\log n}$ and a vector
  $\alpha'$-coloring of~$G[S]$, where $\alpha' =
  \alpha+O(\frac{1}{\log n})$. By Lemma~\ref{L-2}, we can find, in
  polynomial time, an independent set of $G[S]$ of size
  $\Omt(|S|^{f(\alpha')})$. As $|S|=\Omt(n)$, and
  $f(\alpha')=f(\alpha)-O(\frac{1}{\log n})$, we get that the size of
  this independent set, which is also an independent set of~$G$, is
  $\Omt(n^{f(\alpha)})$, as required.
\end{proof}

\section{The coloring tools of Blum}
\label{S-Blum}

Blum \cite{Blum94} makes the following simple observation:

\begin{lemma}[\cite{Blum94}]\label{L-Blum1}
  Let $k\ge 3$ be an integer and let $0<\alpha<1$. If in any
  $k$-colorable graph $G=(V,E)$ on $n$ vertices we can find, in
  polynomial time, at least one of the following:
\begin{enumerate}
\item Two vertices $u,v\in V$ that have the same color under
  some valid $k$-coloring of~$G$ {\rm (Same color)}\label{P-samecolor},
\item An independent set $I\subseteq V$ of size
  $\Omt(n^{1-\alpha})$ {\rm (Large independent set)}\label{P-independent},
\end{enumerate}
 then, we can color every $k$-colorable graph, in polynomial time, using
 $\Ot(n^\alpha)$ colors.
\end{lemma}


If we find one of the objects listed in Lemma~\ref{L-Blum1}
then, following Blum \cite{Blum94}, we say that {\em progress\/} was
made towards coloring the graph using $\Ot(n^\alpha)$ colors. (Blum
\cite{Blum94} describes several other ways of making progress towards
an $\Ot(n^\alpha)$-coloring of the graph which we do not use here.)
We do use the following intricate result which is a small
variant of Corollary~17 of Blum \cite{Blum94}:

\begin{theorem}[\cite{Blum94}]\label{T-Blum}
  Let $G=(V,E)$ be a $k$-colorable graph on $n$ vertices with minimum
  degree $d_{\min}$ in which no two vertices have more than~$s$ common
  neighbors. Then, it is possible to construct, in polynomial time, a
  collection ${\cal T}$ of $\Ot(n)$ subsets of $V$, such that at least
  one $T\in {\cal T}$ satisfies the following two conditions: (i)
  $|T|\ge \Omt(d_{\min}^2/s)$. (ii) $T$ has an independent subset of
  size at least $(\frac{1}{k-1}-O(\frac{1}{\log n}))|T|$.
\end{theorem}

The construction of the collection~$\cT$ is quite simple, though the
proof that at least one of its members satisfies the required
conditions is complicated. For completeness, we present a
self-contained proof of Theorem~\ref{T-Blum} in Appendix~\ref{A-Blum}.

\section{The combined coloring algorithm}
\label{S-new}


We are now able to present the new algorithm for coloring
$k$-colorable graphs using $\Ot(n^{\alpha_k})$ colors, where
$$\begin{array}{c}\alpha_2=0\qu,\qu \alpha_3=\frac{3}{14}\qu,\\[5pt]
\displaystyle  \alpha_k = 1 -
\frac{6}{k+4+3(1-\frac{2}{k})\frac{1}{1-\alpha_{k-2}}}\qu,\qu
\mbox{for $k\ge 4$}\;.\end{array}$$
A description of the algorithm, which we call \combcolor, follows:

\hrulefill

{\bf Algorithm \combcolor:}\\[5pt]
{\bf Input:} A graph $G=(V,E)$ on $n$ vertices and an integer $k\ge
2$. \\
{\bf Output:} An $\Ot(n^{\alpha_k})$ coloring of~$G$, if $G$ is $k$-colorable.

\begin{enumerate}
\item If $k=2$, color the graph, in linear time, using 2 colors.

\item If $k=3$, use the algorithm of Blum and Karger \cite{BlKa97} to
  color the graph using $\Ot(n^{3/14})$ colors.

\item Assume, therefore, that $k\ge 4$. Repeatedly remove from the
  graph~$G$ vertices of degree less than $n^{\alpha_k/(1-2/k)}$. Let
  $U$ be the set of vertices so removed, and let $G[U]$ be the
  subgraph of $G$ induced by~$U$. Let $D$ be the {\em average\/}
  degree of~$G[U]$. It is easy to see that $D\le 2n^{\alpha_k/(1-2/k)}$.
\ignore{If $u,v\in U$, $(u,v)\in E$ and $u$
  was removed before $v$, then direct the edge $(u,v)$ from $u$ to
  $v$. The outdegree of each vertex of $G[U]$ is now at most
  $n^{\alpha_k/(1-2/k)}$.}

\item If $|U|\ge\frac{n}{2}$ then we can use the algorithm of Karger, Motwani and
  Sudan \cite{KaMoSu98} (Theorem~\ref{T-KMS}) to find an
  independent set of $G[U]$ of size $\Omt(n/D^{1-2/k})\ge
  \Omt(n^{1-\alpha_k})$, as $D\le 2n^{\alpha_k/(1-2/k)}$, and we have
  made progress of type~\ref{P-independent}.
\ignore{
color $G[U]$
  using $\Ot(\Delta^{1-2/k})\le \Ot(n^{\alpha_k})$ colors, as
  $\Delta\le n^{\alpha_k/(1-2/k)}$, and we have made progress of
  type~\ref{P-semicolor} (semicoloring).}

\item Otherwise, if $|U|<\frac{n}{2}$, let $W=V{-}U$. Note that
  $|W|\ge\frac{n}{2}$ and that the minimum degree $d_{\min}$ in~$G[W]$
  satisfies $d_{\min}\ge n^{\alpha_k/(1-2/k)}$.

\item For every $u,v\in W$ consider the set $S=N(u)\cap N(v)$. If
  $|S|\ge n^{(1-\alpha_k)/ (1-\alpha_{k-2}) }$, then apply the
  coloring algorithm recursively on $G[S]$ and $k-2$. If $G[S]$ is
  $(k-2)$-colorable, then the algorithm produces a coloring of $G[S]$
  using $\Ot(|S|^{\alpha_{k-2}})$ colors, from which an independent
  set of size $\Omt(|S|^{1-\alpha_{k-2}})\ge \Omt(n^{1-\alpha_k})$ is
  easily extracted, and we have made progress of
  type~\ref{P-independent}. If the coloring returned by the recursive
  call uses more than $\Ot(|S|^{\alpha_{k-2}})$ colors, we can infer
  that $G[S]$ is {\em not\/} $(k-2)$-colorable and thus, $u$ and $v$
  must be assigned the same color under any valid $k$-coloring of $G$,
  as we have made progress of type~\ref{P-samecolor}. 
  \ignore{We are tacitly assuming here that the coloring algorithm
    is deterministic so that it is guaranteed to produce a coloring
    using $\Ot(|S|^{\alpha_{k-2}})$ colors, if $G[S]$ is
    $(k-2)$-colorable. Our algorithm, however, is randomized. There
    are two ways of overcoming this difficulty. The first is to
    derandomize it using the technique of Mahajan and Ramesh
    \cite{MaRa99}. Alternatively, we can simply repeat the whole
    algorithm a sufficient number of times so that the error
    probability is small enough.}

\item Otherwise, we get that $|N(u)\cap N(v)|< n^{(1-\alpha_k)/
    (1-\alpha_{k-2}) }$, for every $u,v\in W$. Also, we know that the
  minimum degree in $G[W]$ is at least $d_{\min}\ge
  n^{\alpha_k/(1-2/k)}$.

\item We can now apply Blum's algorithm \cite{Blum94}
  (Theorem~\ref{T-Blum}), with $d_{\min}\ge n^{\alpha_k/(1-2/k)}$ and $s\le
  n^{(1-\alpha_k)/ (1-\alpha_{k-2}) }$, and obtain a collection $\cT$
  of $\Ot(n)$ subsets of~$W$ such that at least one $T\in \cT$
  satisfies $|T|\ge \Omt(\frac{d^2_{\min}}{s}) \ge
  \Omt\Bigl(n^{\frac{2\alpha_k}{(1-2/k)} -
    \frac{1-\alpha_k}{1-\alpha_{k-2}}}\Bigr)$, and~$T$ contains an
  independent set of size at least $(\frac{1}{k-1}-O(\frac{1}{\log
    n}))|T|$.

\item We now apply the extension of the Alon and Kahale \cite{AlKa98}
  technique (Theorem~\ref{T-Alon-Kahale-extension}) on $G[T]$, for
  each $T\in \cT$. In at least one of these runs we obtain an
  independent set of size $\Omt\Bigl(n^{\bigl(\frac{2\alpha_k}{1-2/k}
    - \frac{1-\alpha_k}{1-\alpha_{k-2}}\bigr)\cdot
    \frac{3}{k}}\Bigr)$. It is easy to check that
  $\bigl(\frac{2\alpha_k}{1-2/k} -
  \frac{1-\alpha_k}{1-\alpha_{k-2}}\bigr)\cdot \frac{3}{k} =
  1-\alpha_k$ (the sequence $\alpha_k$ is defined to satisfy this
  relation), so we have made progress of type~\ref{P-independent}.
\end{enumerate}
\vspace*{-7pt}
\hrulefill

\vspace*{10pt}
The description of \combcolor\ is annotated with a proof that on any
$k$-colorable graph on $n$ vertices it makes progress towards an
$\Ot(n^{\alpha_k})$-coloring of the graph. This, combined with
Lemma~\ref{L-Blum1} gives us the following:

\begin{theorem}\label{T-rc}
  Algorithm \combcolor\ runs in polynomial time and it colors a $k$-colorable
  graph on~$n$ vertices using $\Ot(n^{\alpha_k})$ colors, where
  $\alpha_2=0$, $\alpha_3=\frac{3}{14}$ and $\alpha_k = 1 -
  \frac{6}{k+4+3(1-\frac{2}{k})\frac{1}{1-\alpha_{k-2}}}$,
  for $k\ge 4$.
\end{theorem}

One comment should be made, however. In step~6 of \combcolor\ we are
tacitly assuming that the coloring algorithm is deterministic so
that it is guaranteed to produce a coloring using
$\Ot(|S|^{\alpha_{k-2}})$ colors, if $G[S]$ is $(k-2)$-colorable. Our
algorithm, however, is randomized. There are two ways of overcoming
this difficulty. The first is to derandomize it using the technique of
Mahajan and Ramesh \cite{MaRa99}. Alternatively, we can simply repeat
the whole algorithm a sufficient number of times so that the error
probability is small enough.

\section{Concluding remarks}
\label{S-concl}

We obtained several improved coloring algorithms. It would be
interesting to obtain further improvements. In particular, it would be
interesting to obtain more than logarithmic improvements to the
$\Ot(\Delta^{1-2/k})$ bound of Karger, Motwani and Sudan
\cite{KaMoSu98}, and to see whether better combinations between the
algorithms of Blum \cite{Blum94}, Karger, {\em et al.\/}
\cite{KaMoSu98} and Alon and Kahale \cite{AlKa98} are possible.

\smallskip
Halld\'{o}rsson \cite{Hal93} describes an algorithm for coloring
general graphs using a number of colors which is at most $O(n(\log\log
n)^2/\log^3 n)$ times the minimal number of colors required. His
algorithm is close to being best possible, as it is
known that the chromatic number of general graphs cannot be
approximated, in polynomial time, to within a ratio of $n^{1-\eps}$,
for every $\eps>0$, unless $NP=RP$ (Feige and Killian \cite{FeKi98}).

\smallskip
It is only known, however, that coloring 3-colorable graphs
using 4 colors in NP-hard (Khanna, Linial and Safra \cite{KhLiSa00}
and Guruswami and Khanna \cite{GuKh00}). Obtaining improved hardness
results for coloring 3-colorable graphs is a challenging open problem.

\smallskip 
Another interesting problem is the following: how large can
the chromatic number of vector 3-colorable (or vector $k$-colorable)
graphs be? See Karger {\em et al.\/} \cite{KaMoSu98} for a discussion
of this problem.

\smallskip
Related to the problem of graph coloring is the problem of hypergraph
coloring. See Krivelevich and Sudakov \cite{KrSu98} and Krivelevich,
Nathaniel and Sudakov \cite{KrNaSu01} for the best available results for
this problem.



\begin{thebibliography}{KMS98}

\bibitem[AK98]{AlKa98}
N.~Alon and N.~Kahale.
\newblock Approximating the independence number via the $\theta$-function.
\newblock {\em Mathematical Programming}, 80:253--264, 1998.

\bibitem[BK97]{BlKa97}
A.~Blum and D.~Karger.
\newblock An {$\tilde{O}(n^{3/14})$}-coloring algorithm for 3-colorable graphs.
\newblock {\em Information Processing Letters}, 61:49--53, 1997.

\bibitem[Blu94]{Blum94}
A.~Blum.
\newblock New approximation algorithms for graph coloring.
\newblock {\em Journal of the ACM}, 41:470--516, 1994.

\bibitem[FK98]{FeKi98}
U.~Feige and J.~Kilian.
\newblock Zero knowledge and the chromatic number.
\newblock {\em Journal of Computer and System Sciences}, 57(2):187--199, 1998.

\bibitem[GK00]{GuKh00}
V.~Guruswami and S.~Khanna.
\newblock On the hardness of 4-coloring a 3-colorable graph.
\newblock In {\em Proceedings of the 15th Annual IEEE Conference on
  Computational Complexity, Florence, Italy}, 2000.

\bibitem[GLS93]{GrLoSc93}
M.~Gr{\"{o}}tschel, L.~Lov\'{a}sz, and A.~Schrijver.
\newblock {\em Geometric Algorithms and Combinatorial Optimization}.
\newblock Springer Verlag, 1993.
\newblock Second corrected edition.

\bibitem[Hal93]{Hal93}
M.M. Halld\'{o}rsson.
\newblock A still better performance guarantee for approximate graph coloring.
\newblock {\em Information Processing Letters}, 45:19--23, 1993.

\bibitem[KLS00]{KhLiSa00}
S.~Khanna, N.~Linial, and S.~Safra.
\newblock On the hardness of approximating the chromatic number.
\newblock {\em Combinatorica}, 20:393--415, 2000.

\bibitem[KMS98]{KaMoSu98}
D.~Karger, R.~Motwani, and M.~Sudan.
\newblock Approximate graph coloring by semidefinite programming.
\newblock {\em Journal of the ACM}, 45:246--265, 1998.

\bibitem[KNS01]{KrNaSu01}
M.~Krivelevich, R.~Nathaniel, and B.~Sudakov.
\newblock Approximate coloring of uniform hypergraphs.
\newblock In {\em Proceedings of the 12th Annual ACM-SIAM Symposium on Discrete
  Algorithms, Washington, D.C.}, pages 327--328, 2001.

\bibitem[KS98]{KrSu98}
M.~Krivelevich and B.~Sudakov.
\newblock Approximate coloring of uniform hypergraphs.
\newblock Technical Report 98-31, DIMACS, 1998.

\bibitem[Lov79]{Lovasz79}
L.~Lov{\'{a}}sz.
\newblock On the shannon capacity of a graph.
\newblock {\em IEEE Transactions on Information Theory}, IT-25:1--7, 1979.

\bibitem[MR99]{MaRa99}
S.~Mahajan and H.~Ramesh.
\newblock Derandomizing approximation algorithms based on semidefinite
  programming.
\newblock {\em SIAM Journal on Computing}, 28:1641--1663, 1999.

\bibitem[Wig83]{Wi83}
A.~Wigderson.
\newblock Improving the performance guarantee for approximate graph coloring.
\newblock {\em Journal of the ACM}, 30:729--735, 1983.

\end{thebibliography}

\appendix

\section{Tightness of the refined analysis of Section~\ref{S-KMS}}
\label{A-lower}

To establish the tightness of the analysis presented in
Section~\ref{S-KMS}, we prove the following lemma:

\begin{lemma}
  If $v_1$ and $v_2$ are unit vectors such that $v_1\cdot v_2 =
  -\cos2\beta$, then $$\Pr[v_1 \cdot r \ge c \ {\rm and} \ v_2\cdot r
  \ge c ] = \Omega\left(\frac{1}{c^2}\, \ee^{-\frac{c^2}{2\sin^2\beta}} 
  \right)\;.$$
\end{lemma}

Note, in particular, that for vector 3-colorable graphs we have
$v_1\cdot v_2 = -\frac12$, so $\beta=\frac{\pi}{6}$. As
$\sin\frac{\pi}{6}=\frac12$, we get that $\Pr[v_1 \cdot r \ge c \ {\rm
  and} \ v_2\cdot r \ge c ] = \Omega\left(\frac{1}{c^2}
  \ee^{-2c^2} \right)$, as claimed in Section~\ref{S-KMS}. Also note,
that $\beta<\frac\pi2$ when $v_1\cdot v_2<0$.

\ignore{
We now show that $\Pr[v_1 \cdot r \ge c \ {\rm and} \ v_2\cdot r \ge c
] = \Omega(\frac{1}{c^2}\ee^{-2c^2})$, whenever $v_1\cdot v_2<0$.
Thus, the bound given in Claim~\ref{claim} is asymptotically tight,
and no further improvements can be obtained just by looking at
$\Pr[v_1 \cdot r \ge c \ {\rm and} \ v_2\cdot r \ge c ]$.
}

\begin{figure}[t]
\begin{center}
\setlength{\unitlength}{0.00041667in}
\begingroup\makeatletter\ifx\SetFigFont\undefined%
\gdef\SetFigFont#1#2#3#4#5{%
  \reset@font\fontsize{#1}{#2pt}%
  \fontfamily{#3}\fontseries{#4}\fontshape{#5}%
  \selectfont}%
\fi\endgroup%
{\renewcommand{\dashlinestretch}{30}
\begin{picture}(9849,7239)(0,-10)
\put(6122.833,3582.833){\arc{1909.225}{5.7754}{6.2526}}
\put(2180.108,3564.729){\arc{1215.497}{6.2301}{7.4891}}
\put(2952.000,3679.500){\arc{935.909}{5.9057}{6.3956}}
\put(2037,3612){\blacken\ellipse{150}{150}}
\put(2037,3612){\ellipse{150}{150}}
\put(3087,5412){\blacken\ellipse{80}{80}}
\put(3087,5412){\ellipse{80}{80}}
\put(6222,3597){\blacken\ellipse{100}{100}}
\put(6222,3597){\ellipse{100}{100}}
\put(8442,4902){\blacken\ellipse{100}{100}}
\put(8442,4902){\ellipse{100}{100}}
\path(12,12)(9837,5712)
\path(12,7212)(9837,1512)
\thicklines
\path(2037,3612)(3087,5431)
\path(3039.471,5198.640)(3087.000,5431.000)(2909.561,5273.630)
\path(2037,3612)(3087,1793)
\path(2909.561,1950.370)(3087.000,1793.000)(3039.471,2025.360)
\thinlines
\path(2877,5022)(2487,5247)(2712,5637)
\path(2712,1572)(2487,1962)(2877,2187)
\path(2112,3612)(9837,3612)
\path(2082,3582)(8472,4902)(8472,4917)
\put(2187,2562){\makebox(0,0)[b]{\smash{{{\SetFigFont{10}{12.0}{\rmdefault}{\mddefault}{\updefault}$cv_2$}}}}}
\put(1737,3537){\makebox(0,0)[b]{\smash{{{\SetFigFont{10}{12.0}{\rmdefault}{\mddefault}{\updefault}$O$}}}}}
\put(2187,4662){\makebox(0,0)[b]{\smash{{{\SetFigFont{10}{12.0}{\rmdefault}{\mddefault}{\updefault}$cv_1$}}}}}
\put(7362,3912){\makebox(0,0)[lb]{\smash{{{\SetFigFont{10}{12.0}{\rmdefault}{\mddefault}{\updefault}$\beta$}}}}}
\put(2847,3087){\makebox(0,0)[lb]{\smash{{{\SetFigFont{10}{12.0}{\rmdefault}{\mddefault}{\updefault}$\frac{\pi}{2}-\beta$}}}}}
\put(5697,4587){\makebox(0,0)[lb]{\smash{{{\SetFigFont{10}{12.0}{\rmdefault}{\mddefault}{\updefault}$r$}}}}}
\put(3762,3657){\makebox(0,0)[lb]{\smash{{{\SetFigFont{10}{12.0}{\rmdefault}{\mddefault}{\updefault}$\theta(r)$}}}}}
\put(5007,3657){\makebox(0,0)[lb]{\smash{{{\SetFigFont{10}{12.0}{\rmdefault}{\mddefault}{\updefault}$R$}}}}}
\end{picture}
}
\end{center}
\caption{Lower bounding $\Pr[v_1 \cdot r \ge c \ {\rm and} \ v_2\cdot r
  \ge c ]$}
\label{F-wedge}
\end{figure}
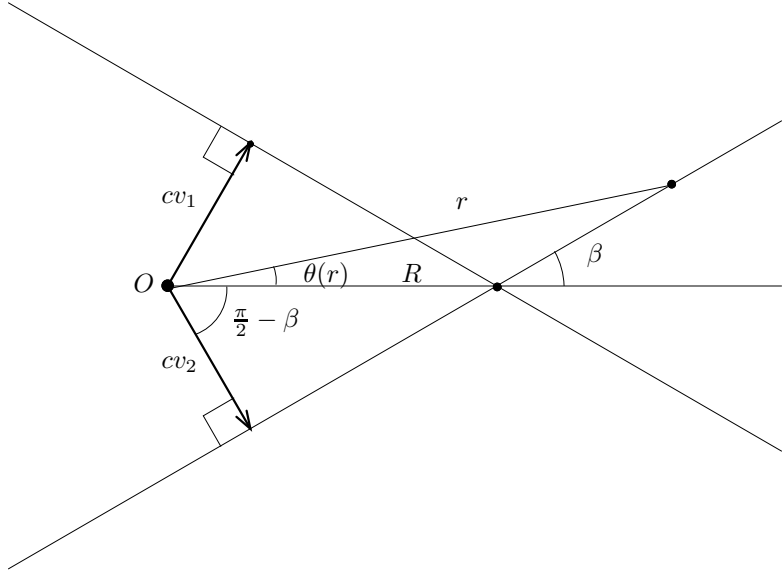

\begin{proof}
Let $P(\beta) = \Pr[v_1 \cdot r \ge c \ {\rm and} \ v_2\cdot r
\ge c ]$. Consulting
Figure~\ref{F-wedge}, we see that
$$P(\beta)\eq \int\!\!\int_{(x,y)\in W(\beta)}
\phi(x)\phi(y)\,dx\,dy \eq \frac{1}{2\pi} \int\!\!\int_{(x,y)\in
  W(\beta)} \ee^{-(x^2+y^2)/2}\,dx\,dy\;,$$
where
$$W(\beta)\eq \{\, (x,y)\in\reals^2 \mid -(x-R)\tan\beta \le y \le
(x-R)\tan\beta\,\}\;,$$
and 
$$R\eq R(\beta)\eq \frac{c}{\cos(\frac{\pi}{2}-\beta)}=\frac{c}{\sin\beta}\;.$$
Moving to polar coordinates, we get that
$$P(\beta) \eq \int\!\!\int_{(x,y)\in W'(\beta)}
r\,\ee^{-r^2/2}\,dr\,d\theta \eq
\frac{1}{\pi}\int_{R}^\infty\left[ \int_0^{\theta(r)}
  r\,\ee^{-r^2/2}\,d\theta \right]\,dr \eq \frac{1}{\pi}
\int_{R}^\infty \theta(r)\,r\,\ee^{-r^2/2}\,dr\;,$$
where $W'(\beta)$ is the region $W(\beta)$ expressed in polar coordinates.
Using the sine theorem, we get that
$$   \frac{\sin(\pi - \beta)}{r} = \frac{\sin(\beta - \theta(r))}{R}\;,  $$
and thus 
$$\theta(r) = \beta - \arcsin \frac{c}{r}\;.$$
Putting all this together, we get that
$$ P(\beta)\eq \frac{1}{\pi} \int^{\infty}_{R}(\beta - \arcsin \frac{c}
       {r})\, r\, e^{-{r^2}/{2}}\,dr\;.$$
Next, we change the variable of integration. Let $r=\frac{c}{\sin
  t}$, so that
$dr=-\frac{c \cos{t}}{\sin^2t}\,dt$. We get that
\begin{eqnarray*}
 P(\beta) &=& 
    \frac{1}{\pi} \int^{0}_{\beta} (\beta - t) \left(\frac{c}{\sin t}\right)
\left(\ee^{-\frac{c^2}{2\sin^2 t}}\right)
 (-\frac{c \cos t}{\sin^2t})\,dt\\ 
 &=& \frac{1}{\pi} \int^{\beta}_{0} (\beta - t)
    \left(\ee^{-\frac{c^2}{2\sin^2t}}\right) \left(\frac{c^2 \cos
    t}{\sin^3t}\right)\,dt  \\
& = & \frac{1}{\pi} \int^{\beta}_{0} (\beta - t)
    \left[\ee^{-\frac{c^2}{2\sin^2t}}\right]'\,dt  
\end{eqnarray*}
Using integration by parts we finally get the concise formula:
$$ P(\beta) \eq \frac{1}{\pi}\int^{\beta}_0 \ee^{-\frac{c^2}{2\sin^2t}}\,dt\;.$$

Let us now consider the integral
\begin{displaymath}
 Q(\beta) = \frac{1}{\pi}\int^{\beta}_0 
\ee^{-\frac{c^2}{2\sin^2t}}(2\sin^2t+\tan^2t)\,dt\;.
\end{displaymath}
Since $2\sin^2t+\tan^2t$ is an increasing function for $0\le
 t<\frac{\pi}{2}$, we get that
$$Q(\beta)\;\le\; A(\beta)P(\beta) \quad {\rm where} \quad
A(\beta)\eq 2\sin^2\beta+\tan^2\beta\;.$$
On the other hand, by integrating by parts, we get that
\begin{eqnarray*}
Q(\beta) &=&  \frac{1}{\pi}\int^{\beta}_0 
\ee^{-\frac{c^2}{2\sin^2t}} \left[\frac{\sin^3t}{\cos t}\right]'\,dt\\
&=&
\left.
  \frac{1}{\pi}\ee^{-\frac{c^2}{2\sin^2t}}\left(\frac{\sin^3t}{\cos
      t}\right) \right|_0^\beta -
\frac{1}{\pi}\int^{\beta}_0 \ee^{-\frac{c^2}{2\sin^2t}} 
 \left(\frac{c^2\cos t}{\sin^3t}\right)
 \left(\frac{\sin^3t}{\cos t}\right)\,dt \\
&=&  \frac{1}{\pi}\ee^{-\frac{c^2}{2\sin^2\beta}}\left(\frac{\sin^3\beta}{\cos
      \beta}\right) - c^2\, P(\beta)\;.
\end{eqnarray*}
Letting $B(\beta)= \frac{1}{\pi}\frac{\sin^3\beta}{\cos
  \beta}$, we get that
$$B(\beta)\ee^{-\frac{c^2}{2\sin^2\beta}} - c^2 P(\beta) \eq
Q(\beta) \;\le\; A(\beta) P(\beta)\;,$$
and thus
$$P(\beta)\;\ge\; \frac{B(\beta)}{c^2+A(\beta)}\,
\ee^{-\frac{c^2}{2\sin^2\beta}} \eq \Omega\left(\frac{1}{c^2}\,
\ee^{-\frac{c^2}{2\sin^2\beta}} \right)\;,$$
as claimed.
\end{proof}

\section{Proof of Theorem~\ref{T-Blum}}
\label{A-Blum}

We begin by introducing some notation.  For a vertex~$v$, let $d(v)$ be the
degree of~$v$, and $N(v)$ be the set of neighbors of~$v$. For a set
$S\subseteq V$, let $D(S) = \sum_{v \in S} d(v)$, let
$d_S(v)=|N(v)\cap S|$ be the number of neighbors of~$v$ in~$S$, and
let $N(S)=\cup_{v\in S}N(v)$ be the set of neighbors of~$S$. For
another set $T\subseteq V$, let $D_T(S) = \sum_{v \in S} d_T(v)$.
Clearly, $D_T(S) = D_S(T)$.

We consider a certain $k$-coloring of the graph,
i.e., a partition of the graph into $k$ disjoint independent
sets $S_1,S_2,\ldots, S_k$, and we assume, without loss of
generality, that $D(S_1) \geq D(S_i)$ for $i=1\ldots,k$. We
call the vertices from $S_1$ {\em red} vertices, and let $R
= S_1$. By the choice of~$R$, we have that
$D_R(V- R) = D(R) \geq D(V - R)/(k-1)$. We shall use the following
simple claim:

\begin{claim}\label{pigeon}
Let $x_1,\ldots,x_n\ge 0$ and $y_1,\ldots,y_n\ge 0$ be
such that $\sum_{i=1}^n x_i = \alpha n$ and $\sum_{i=1}^n
x_i \ge \beta \sum_{i=1}^n y_i$. Then, for every $\delta > 0$
there is at least one index $1 \leq i \leq n$ which satisfies 
$$x_i \;\geq\; \delta\alpha \quad , \quad x_i \;\geq\;
(1-\delta)\beta y_i\;.$$
\end{claim}
\begin{proof}
Let $I = \{\,1\le i\le n \mid x_i \geq \delta\alpha\,\}$. It
is easy to see that 
$$\sum_{i \in I} x_i \;\geq\; (1-\delta)\sum_{i = 1}^n x_i \;\geq\;
(1-\delta)\beta \sum_{i \in I} y_i,$$
and therefore, there is at least one $i \in I$, such that
$x_i \geq (1-\delta)\beta y_i$, and the claim follows.
\end{proof}

We now show that for at least one {\em red} vertex~$v$, there is a large
subset~$S$ of $N(v)$, such that the set $N(S)$ contains
relatively many {\em red} vertices, and $|S| =
\Omt(d_{min})$. As each vertex of $N(S)$ has at most $s$
common neighbors with~$v$, we get that $|N(S)| \geq
|S|d_{min}/s$, and thus the theorem would follow. We begin with
the following lemma: 

\begin{lemma}\label{regular-Blum}
  Let $U \subseteq V - R$ be such that $d \leq d(v) < d(1+\delta)$,
  for every $v\in U$. (In other words, all the vertices of~$U$ are of
  roughly the same degree.)  If $D_R(U) \geq \lambda D(U)$, for some
  $\lambda>0$, then there is a {\em red} vertex~$v$ such that
\begin{equation}\label{red-edges}
 D_R(N(v) \cap U) \geq (1-\delta)\lambda D(N(v) \cap U).
 \end{equation}
\end{lemma}

\begin{proof}
Assume, for contradiction, that Equation~(\ref{red-edges}) does
not hold for any red vertex. If we sum up over all red
vertices, we get that
\begin{equation}\label{contr1}
 \sum_{v \in R} D_R(N(v) \cap U) < (1-\delta)\lambda\sum_{v
 \in R}D(N(v) \cap U)\;.
\end{equation}
Now,
$$\begin{array}{c}
\displaystyle
\sum_{v \in R} D_R(N(v) \cap U) \eq  \sum_{v \in R} \sum_{u \in
   N(v) \cap U} d_R(u) 
 \eq  \sum_{u \in U} d_R(u)^2\;, \\[15pt]
\displaystyle
\sum_{v \in R} D(N(v) \cap U) \eq  \sum_{v \in R} \sum_{u \in
   N(v) \cap U} d(u) 
 \eq  \sum_{u \in U} d_R(u)d(u) 
 \;<\; d (1+\delta)\sum_{u \in U} d_R(u)\;. \\
\end{array}$$
Combining this with~(\ref{contr1}) and the Cauchy-Schwartz
inequality, we get that
$$ D_R(U) \eq \sum_{u \in U} d_R(u) 
\leq  \frac{|U|\sum_{u \in U}  d_R(u)^2}{\sum_{u \in U} d_R(u)} 
 \;\leq\; \lambda d(1-\delta)(1+\delta)|U|  \;<\; \lambda D(U)\;,$$
a contradiction.
\end{proof}

We are now ready to prove Theorem~\ref{T-Blum}.

\begin{proof}\hspace*{-0.2cm}
  {\bf (of Theorem~\ref{T-Blum})}\quad Let $\delta = \frac{1}{\log
    n}$, and let $I_j = \{v \in V - R \mid (1+\delta)^j \leq d(v) <
  (1+\delta)^{j+1} \}$, for $1\le j\le \log_{1+\delta} n$ . By
  Claim~\ref{pigeon}, with $x_j = D_R(I_j)$ and $y_j = D(I_j)$, at
  least one such set $I_j$ satisfies
\begin{equation}\label{big-DR}
D_R (I_j) \;\geq\; \delta \frac{D_R(V - R)}{\log_{1+\delta}n}
 \quad,\quad
D_R(I_j) \;\geq\; (1-\delta)\frac{D(I_j)}{k-1}
\;.
\end{equation}
 We now remove from the graph all the
red vertices~$v\in R$,
for which $N(v) \cap I_j$ is small. More formally, we remove all
vertices  $v \in R$ for which $d_{I_j}(v) < \delta^2
 d_{\min}/\log_{1+\delta}  n$. We let $R'$ be the remaining set of red
 vertices. 
 It is easy to see, by~(\ref{big-DR}), that in the
 remaining graph we have $D_{R'}(I_j) \geq (1-\delta)D_R(I_j)$,
 and thus, we can apply Lemma~\ref{regular-Blum}, with $U =
 I_j$, $R = R'$, $\lambda = (1-\delta)^2/(k-1)$, and we get a set $S = N(v)
 \cap I_j$, such that $|S| = \Omt(d_{\min})$ and $D_R(S) \geq
(1-\delta)^3\frac{D(S)}{k-1}$.

For every $u \in N(S)$, we know that $|N(u) \cap S| \leq
   s$, and therefore $|N(S)| \geq D(S)/s = \Omt(d_{\min}^2/s)$. 
If  all vertices in $N(S)$ have the same degree into~$S$,
then clearly $|N(S) \cap R| \geq (1-\delta)^3|N(S)|/(k-1)$,
   and we are done. We therefore partition the 
vertices of $N(S)$ into sets of vertices with roughly the
same degree into~$S$, $N_i(S) = \{ u \in N(S) \mid (1+
\delta)^i \leq d_S(u) < (1+ \delta)^{i+1}\}$. By
Claim~\ref{pigeon}, with $x_i = D_{R \cap N_i(S)}(S)$ and
$y_i = D_{N_i(S)}(S)$, there is at least one set $N_i(S)$,
such that 
$$ D_{N_i(S)}(S) \;\geq\; D_{R \cap N_i(S)}(S) \;=\; \Omt( d_{\min}^2)
\quad,\quad  D_{R\cap
   N_i(S)}(S) \;\geq\; (1-\delta)^4 \frac{D_{N_i(S)}(S)}{k-1}.$$
For every $u \in N_i(S)$, we know that $|N(u) \cap S| \geq
   s$, and therefore $|N_i(S)| \geq D_{N_i(S)}(S)/s =
   \Omt(d_{\min}^2/s)$. In $N_i(S)$, the degrees into~$S$ 
   are roughly the same, and thus, $|N_i(S) \cap R| \geq
   (1-\delta)^5|N_i(S)|/(k-1)$.  
   
   Thus, we proved that in the collection ${\cal T} = \{T_{ij} =
   N_i(N(v) \cap I_j)\}$, whose size is $O(n\log^2_{1+\delta}n)$,
   there is at least one set $T \in {\cal T}$ that satisfies the
   required properties.
\end{proof}

\end{document}